\documentstyle{article}
\begin{document}

\overfullrule=0pt
\baselineskip=18pt

\title{Scattering from Spatially Localized Chaotic and Disordered Systems}
\author{L.E. Reichl and G. Akguc\\
Center for Studies in Statistical Mechanics and Complex Systems\\
The University of Texas at Austin\\
Austin, Texas 78712}

\date{\today}

\maketitle
\begin{abstract}

A version of scattering theory that was developed many years ago to treat
nuclear scattering processes, has provided a powerful tool to study
universality in scattering processes involving open quantum systems with
underlying
classically chaotic dynamics. Recently, it has been used to make random
matrix theory predictions concerning the statistical properties of
scattering resonances in mesoscopic electron waveguides and electromagnetic
waveguides. We provide a simple derivation of this scattering theory and
we compare its predictions to those obtained from an exactly solvable
scattering model; and we use it to study the scattering of a particle wave
from a random potential.  This method may prove useful in distinguishing
the effects of chaos from the effects of disorder in real scattering
processes.

\end{abstract}

\bigskip

\section{Introduction}
Interest in the dynamical properties of open quantum systems at mesoscopic and
atomic scales has lead to a rebirth of a form of scattering theory which was
originally developed to deal with very complicated nuclear collision processes.
The origin of the  scattering theory that we consider here came
from the recognition that a collision between two nucleons,  one of which
is very heavy, can lead to the creation of an unstable, but very long-lived,
compound nucleus which
eventually decays. In the late 1930's, Kapur and Peierls \cite{kn:1}
 used this fact to formulate a
nonperturbative approach to scattering theory in which the compound nucleus
was viewed as
a stable object which was made unstable by weak coupling to the continuum.
In the late 1940's,
Wigner and Eisenbud \cite{kn:2} developed
an alternate version of the  Kapur-Peierls theory which lead to the very
widely used
R-matrix approach to scattering theory \cite{kn:3}. The idea behind both these
theories is to decompose  configuration space into an internal region
(reaction region) and
an external asymptotic scattering region.
As we shall see, this approach can be made clean and
rigorous if there is an abrupt (in configuration space) transition between
the scattering
region and the asymptotic region. The internal region can be modeled
in terms of a
complete set of states with fixed boundary conditions on the surface of the
internal region.
The internal states can then be coupled to the external asymptotic
states. Bloch \cite{kn:4} and
 Feshbach \cite{kn:5} both showed that a consistent theory requires that the
coupling between the
internal and external regions must be
singular.
In the 1960's, Weidenmuller developed a phenomenological Hamiltonian
approach to
nuclear scattering based on this picture \cite{kn:6}.
 The Hamiltonian for the interior region was based
on the shell model of the nucleus. The Hamiltonian for the exterior
scattering region
described the asymptotic states. The strength of the coupling between the
interior and exterior
regions was an unknown input parameter.

The Weidenmuller approach to scattering theory created a framework with
which to study the
manifestations of chaos in the scattering properties of nuclear systems, as
well as
mesoscopic and atomic systems.  It is now well known that
chaos manifests itself in bounded quantum systems by affecting
 the statistical distribution of spacings
between energy levels \cite{kn:7}. For open quantum systems,
in regimes where scattering resonances are not strongly overlapping,
underlying chaos affects
the statistical distribution of resonance spacings and resonance
widths \cite{kn:8}.  If the
Hamiltonian of the
interior region is chosen to be a random matrix Hamiltonian, then
predictions can be made as
to the statistical properties of resonance spacings and resonance widths.
These predictions can
then be compared to nuclear  scattering resonances, or to resonances in
electron waveguides
[\cite{kn:9}], \cite{kn:12},and [\cite{kn:10}] and
electro-magnetic resonators
[\cite{kn:1}] and \cite{kn:13} whose cavities have been
constructed to yield classically chaotic behavior.
The  agreement between random matrix theory predictions and numerical and
laboratory
experiments is quite good.

The Weidenmuller Hamiltonian is phenomenological and does not have
information about the
strength of the interface coupling. The random matrix theories
treat the interface coupling
as an input parameter. In a real dynamical system, however, one needs a way
to rigorously
determine the coupling at the interface of the internal and external
regions. A very useful
way to fix the coupling at the interface was provided by Pavlov
\cite{kn:11}.
The idea of Pavlov was that the surface coupling could be fixed by the
requirement that the
total Hamiltonian (including interior and exterior regions) be Hermitian.

In this paper, we use a simple textbook  scattering problem to illustrate
and clarify many issues concerning this
approach to scattering,  and we then use it
to study the effect of disorder on the scattering process.

In Section (2), we describe the ``textbook" scattering problem and obtain exact
expressions for the reaction function, the scattering function, and the
Wigner delay times.
In Section (3), we develop the Hamiltonian for interior and exterior
regions of the
scattering system.
In Section (4) we derive
the Hermiticity condition. In Section (5), we derive the reaction function. In
Section (6),
we derive the scattering function and locate resonance energies for the case
of a smooth scattering potential. In Section (7), we use this theory
to study the scattering of a particle wave from a random scattering
potential. Finally,  in Section (8), we make some concluding remarks.

\bigskip
\bigskip
\section{Exact Solution}

We will first consider the scattering of a particle of mass, $m$, due to the
potential shown in Fig. (1). The
quantum particle enters from the right with energy E and is reflected back
to the left  by an
infinitely hard
wall located at $x=0$. A barrier of height, $V_0$, is located $0<x<a$.
The Schroedinger equation, which describes propagation of a particle wave,
$\Psi(x,t)$, for all times, $t$, is given by
$$i{\hbar}{{\partial}{\Psi}(x,t)\over {\partial}t}=
-{{\hbar}^2\over 2m}{{\partial}^2{\Psi}(x,t)\over
{\partial}x^2}+V(x){\Psi}(x,t),\eqno(2.1)$$
where ${\hbar}$ is Planck's constant and the potential,
$V(x)={\infty}$ for $(-\infty<x{\leq}0)$,
$V(x)=V_0$ for $(0<x{\leq}a)$, and
$V(x)=0$ for $(a<x<\infty)$. Since the potential is infinite for
$-\infty<x<0$, no
wavefunction can exist in that region of space.

The Schroedinger equation for energy eigenstates ${\Phi}_E^I(x)$,
for Region I ($0<x<a$) in Fig. (1), is
$$-{{\hbar}^2\over 2m}{d^2{\Phi}^I_E(x)\over
dx^2}+V_0{\Phi}_E(x)=E{\Phi}^I_E(x),\eqno(2.1)$$
Energy eigenstates in Region I have the form
$${\Phi}^I_E(x)=A{\sin}(k'x)\eqno(2.2)$$
where $k'=\sqrt{{2m\over {\hbar}^2}(E-V_0)}$ and $A$ is a normalization
constant.
The Schroedinger equation for energy eigenstates ${\Phi}_E^{II}(x)$,
for Region II ($a<x<-\infty$) in Fig. 1, is
$$-{{\hbar}^2\over 2m}{d^2{\Phi}^{II}_E(x)\over
dx^2}=E{\Phi}^{II}_E(x),\eqno(2.3)$$
Energy eigenstates in Region II have the form
$${\Phi}^{II}_E(x)=B({\rm e}^{-ikx}-S(E){\rm e}^{ikx}),\eqno(2.4)$$
where $k=\sqrt{{2m\over {\hbar}^2}E}$. The first term describes the
incoming part
of the energy eigenstate and the second term describes the outgoing part of that
state. The constant, $B$, is a normalization constant and $S(E)$ is the
{\it scattering function}. When $V_0=0$, then $S(E)=1$ and the incoming wave is
reflected from the wall at $x=0$ with an overall phase shift of $\pi$ since
we can
rewrite the minus sign as ${\rm e}^{i\pi}$. Thus, $S(E)$, contains information
about the phase shift of the scattered wave due to the potential barrier, $V_0$,
in the region, $0<x<a$.

We can determine how the two parts of the energy eigenstate are connected
by the
condition that the wavefunction and the slope of the wavefunction must
be continuous at $x=a$. These two conditions can be combined into a statement
that the "logarithmic derivative" of the wavefunction
must be continuous at the interface. Thus we
must have
$${1\over R(E)}{\equiv}{a\over {\Phi}^I_E(a)}{d{\Phi}^I_E\over dx}{\bigg|}_a=
{a\over {\Phi}^{II}_E(a)}{d{\Phi}^{II}_E\over dx}{\bigg|}_a\eqno(2.5)$$
at the interface. The function, $R(E)$, is called the {\it reaction function}
and it will play an important role in the following sections. The S-matrix
is then given by
$$S(E)={\rm e}^{-2ika}{(1+ikaR(E))\over (1-ikaR(E))}.\eqno(2.6)$$
where the reaction function,
$$R(E)={{\tan}(k'a)\over k'a}.\eqno(2.7)$$
Since the
scattering function has unit magnitude we can also write it
$$S(E)={\rm e}^{i{\theta}(E)}={\rm e}^{-2ika}
{\rm e}^{i{\theta}'(E)},\eqno(2.8)$$
where ${\theta}(E)$ is the phase shift of the scattered wave relative to
the case for which $V_0=0$.
It is straightforward to show that
$${\rm tan}({\theta}')={2kaR(E)\over 1-k^2a^2R(E)^2}~~~{\rm and}~~~
{\rm tan}({{\theta}'\over 2})=kaR(E).\eqno(2.9)$$
The time delay of the scattered particle due to its interaction with the
barrier, $V_0$, is given by
$${\tau}{\equiv}{\hbar}{d\theta\over dE}{\big|}_{k=k_0}.\eqno(2.10)$$
This time delay is called the {\it Wigner delay time} [\cite{bohm}].

For the scattering system we consider here, the phase angle,
${\theta}'(E)$ is given by
$${\theta}'(E)=2{\rm arctan}{\bigl[}{{\beta}\over {\beta}'}
{\rm tan}({\beta}'){\bigr]},\eqno(2.11)$$
where ${\beta}=ka$ and ${\beta}'=k'a$.
The Wigner delay time is given by
$${\tau}(E)={\hbar}{d{\theta}(E)\over dE}$$
$$~~~={2ma^2\over {\hbar}^2}
{\biggl(}{1\over {\beta}'^2+{\beta}^2{\rm tan}({\beta}')^2}
{\biggl[}{\biggl(}{{\beta}'\over {\beta}}-{{\beta}'\over {\beta}}{\biggr)}
{\tan}({\beta}')+{\beta}{\sec}({\beta}')^2{\biggr]}
{\biggr)}-{2ma^2\over {\hbar}{\beta}}.\eqno(2.12)$$
 In Fig. (2.a)
we show the phase angle, ${\theta}(E)$, as a function of energy, and
in Fig. (2.b), we show the Wigner delay time for parameter values,
${\hbar}=1$, $m=1$, $a=1$, and $V_0=10$. The peaks in Fig. (2.b) occur at
values of the energy where the particle wave resonates with the barrier
region. The fact that the Wigner delay time goes negative for low energies
means that when $V_0{\neq}0$, the
particle can be reflected earlier by the barrier, $V_0$,
than it would be if $V_0=0$.

In Fig. (3), we show that exact energy eigenstates as a function of
position for a
range of energies.  At resonant energies, the energy eigenfunctions have
enhanced
probability in the scattering region.  These resonances are also often called
{\it quasi-bound states} because they are associated with complex energy poles
of the scattering function. In Fig. (4), we plot those complex energy poles
of $S(E)$ which give rise to the first three resonance peaks in the Wigner
delay time. It is important to note that for scattering systems with higher
space dimension, complex energy poles can have a profound effect on the
scattering properties. For example, in two dimensional electron wave guides
they can completely block transmission in some channels [\cite{na}],
[\cite{ree}].

\bigskip
\bigskip
\section{Scattering Hamiltonian}

Most of the interesting information regarding scattering events is
contained in the
energy dependence of the scattering phase shifts and in the location of
quasi-bound
state poles.  However, we generally cannot find those quantities exactly as
we did in
the previous section. There are a number of techniques for finding them if
the scattering potential is weak, but not if the
scattering potential is strong. In subsequent sections, we describe
an approach to scattering theory which can deal with strong interactions,
provided
they are confined to a localized region of space. As we mentioned in Sect. (1),
this approach to scattering theory was originally developed to deal with
nuclear
scattering processes, but recently has provided an important tool for studying
universal properties of scattering processes induced by underlying chaos.
We shall
use this alternate approach to scattering  theory to study the system
considered in
Sect. (2), and a similar system with a random scattering potential. In this
way, we can compare its predictions to the exact results which
were obtained in Sect. (2).

Let ${\hat x}$ denote the position
operator and let $|x{\rangle}$ denote its eigenstates, so that ${\hat
x}|x{\rangle}=x|x{\rangle}$.
The position
eigenstates satisfy a completeness relation,
${\int_{-\infty}^{\infty}}dx~|x{\rangle}{\langle}x|={\hat 1}$,
where ${\hat 1}$ is the unit operator, and they are delta normalized,
${\int_{-\infty}^{\infty}}dx~{\langle}x|x'{\rangle}={\delta}(x-x')$. Since
the potential energy
is infinite for $x<0$, all states will be zero in that region. We can
divide the completeness
relation into three parts, ${\hat
N}={\int_{-\infty}^{0}}dx~|x{\rangle}{\langle}x|$,
${\hat Q}={\int_{0}^{a}}dx~|x{\rangle}{\langle}x|$, and
${\hat P}={\int_{a}^{\infty}}dx~|x{\rangle}{\langle}x|$, so that ${\hat
N}+{\hat Q}+{\hat
P}={\hat 1}$. However, the operator
${\hat N}$ acting on any state
gives zero because the potential energy is infinite in that region.
Therefore, we can remove
${\hat N}$ from the completeness relation without changing
our final results. Thus, from now on we will write the completeness relation
as ${\hat
Q}+{\hat P}={\hat 1}$.

The operators, ${\hat Q}$  and ${\hat P}$ , are projection operators. They
have the property
that ${\hat Q}={\hat Q}^2$, ${\hat P}={\hat P}^2$, and ${\hat Q}{\hat
P}={\hat P}{\hat Q}=0$.
This is easily checked by explicit calculation.
The operator ${\hat Q}$ projects any state or operator onto the interval,
$0{\leq}x{\leq}a$,
while the operator ${\hat P}$ projects any state or operator
onto the interval, $a{\leq}x{\leq}\infty$. In other words, if the state,
$|{\Psi}{\rangle}$ has spatial dependence,
${\Psi}(x){\equiv}{\langle}x|{\Psi}{\rangle}$, over the interval
$(0<x<\infty)$,
then the state ${\langle}x|{\hat Q}|{\Psi}{\rangle}={\Psi}(x)$ for
$(0<x<a)$ and the
state ${\langle}x|{\hat P}|{\Psi}{\rangle}={\Psi}(x)$ for $(a<x<\infty)$.

Inside the Region I, $(0<x<a)$ in Fig.(1), we define a Hamiltonian,
$${\hat H}_{QQ}{\equiv}{\hat Q}{\bigl(}
{1\over 2m}{\hat p}^2+V_0{\bigr)}{\hat Q},\eqno(3.1)$$
where ${\hat p}$ is the
momentum operator and
$m$ is the mass of the particle. The Hamiltonian, ${\hat H}_{QQ}$, is Hermitian
and therefore it will have a complete, orthonormal set of eigenstates which we
denote as
${\hat Q}|{\phi}_j{\rangle}$.  We can write the eigenvalue problem in the
region, $0<x<a$, as
${\hat H}_{QQ}{\hat Q}|{\phi}_j{\rangle}={\lambda}_j{\hat
Q}|{\phi}_j{\rangle}$, where
${\lambda}_j$ is the $j^{th}$ energy eigenvalue of ${\hat H}_{QQ}$ and
$j=1,2,...{\infty}$.
Because there is an infinitely hard wall at $x=0$, the eigenstates
${\phi}_j(x){\equiv}{\langle}x|{\hat Q}|{\phi}_j{\rangle}$ must be zero at
$x=0$. We have some freedom
in choosing the boundary condition
at $x=a$ and we will do that later. The completeness of the states,
${\hat Q}|{\phi}_j{\rangle}$, allows us to
write the completeness relation,
${\sum_j}{\hat Q}|{\phi}_j{\rangle}{\langle}{\phi}_j|{\hat Q}={\hat Q}$.
Orthonormality requires
that ${\langle}{\phi}_j|{\hat Q}|{\phi}_{j'}{\rangle}={\delta}_{j,j'}$.

The Hamiltonian for the region, $a<x<\infty$, can be written
$${\hat H}_{PP}{\equiv}{\hat P}{\bigl(}
{1\over 2m}{\hat p}^2{\bigr)}{\hat P}.\eqno(3.2)$$
Its eigenvalues are
continuous and have range, $(0{\leq}E{\leq}\infty)$. The eigenvector of
${\hat H}_{PP}$, with eigenvalue, $E$, will be denoted ${\hat
P}|E{\rangle}$. The eigenvalue equation then reads,
${\hat H}_{PP}{\hat P}|E{\rangle}=E{\hat P}|E{\rangle}$.

Any state, $|{\Psi}{\rangle}$, can be decomposed into its contributions from
the two disjoint
regions of configuration space as
$|{\Psi}{\rangle}={\hat Q}|{\Psi}{\rangle}+{\hat P}|{\Psi}{\rangle}$. We
can expand ${\hat Q}|{\Psi}{\rangle}$ in terms of the complete set of states,
${\hat Q}|{\phi}_j{\rangle}$ and we obtain,
$$ |{\Psi}{\rangle}={\sum_j}{\beta}_j{\hat Q}|{\phi}_j{\rangle}
+{\hat P}|{\Psi}{\rangle},\eqno(3.3)$$
where ${\beta}_j={\langle}{\phi}_j|{\hat Q}|{\Psi}{\rangle}$.
The function ${\langle}x|{\hat Q}|{\Psi}{\rangle}$
must be equal to the function, ${\langle}x|{\hat P}|{\Psi}{\rangle}$,
at the interface, $x=a$.  In addition, the slopes of these two functions
must be equal at the interface.

We couple the two regions of configuration space, at their interface,
with the singular operator, ${\hat V}=C{\delta}({\hat x}-a){\hat p}$. The
coupling
constant,
$C$, will be determined later.
Then
$${\hat H}_{QP}={\hat
Q}{\hat V}{\hat
P}=C{\int_0^a}dx_1~{\int_a^{\infty}}dx_0~|x_1{\rangle}
 {\delta}(x_1-x_0)
 {\delta}(x_0-a){d\over dx_0}{\langle}x_0|,\eqno(3.4)$$
and
$${\hat H}_{PQ}={\hat
P}{\hat V}{\hat
Q}=C{\int_a^{\infty}}dx_0~{\int_0^a}dx_1~|x_0{\rangle}
{\delta}(x_0-x_1)
 {\delta}(x_1-a){d\over dx_1}{\langle}x_1|.\eqno(3.5)$$
It is useful to remember that ${\int_0^a}dx~{\delta}(x-a)={1\over 2}$,
${\int_0^a}dx~{\delta}(x-x_0)=1$ if $0<x_0<a$, and
${\int_0^a}dx~{\delta}(x-x_0)=0$ if $a<x_0$.
Note also that
$${\hat H}_{QQ}={\int_0^a}dx~|x{\rangle}
{\biggl(}{-{\hbar}^2\over 2m}{d^2\over dx^2}+V_0{\biggr)}
{\langle}x|\eqno(3.6)$$
and
$${\hat H}_{PP}={\int_a^{\infty}}dx~|x{\rangle}
{\biggl(}{-{\hbar}^2\over 2m}{d^2\over dx^2}{\biggr)}
{\langle}x|\eqno(3.7)$$

The total Hamiltonian of the system can be written
$${\hat H}={\hat H}_{QQ}+{\hat H}_{PP}+{\hat H}_{QP}+{\hat H}_{PQ}.\eqno(3.8)$$
The energy eigenstates, $|E{\rangle}$, satisfy the eigenvalue
equation
${\hat H}|E{\rangle}=E|E{\rangle}$. The energy eigenstates can be
decomposed into
their contributions
from the two regions of configuration space, so that
$$|E{\rangle}={\sum_j}{\gamma}_j{\hat Q}|{\phi}_j{\rangle}
+{\hat P}|E{\rangle},\eqno(3.9)$$
where ${\gamma}_j={\langle}{\phi}_j|{\hat Q}|E{\rangle}$. The eigenvalue
equation then
takes the form

$$\pmatrix{{\hat H}_{QQ}&0&\ldots&{\hat H}_{QP}\cr
0&{\hat H}_{QQ}&\ldots&{\hat H}_{QP}\cr
\vdots&\vdots&\ddots&\vdots\cr
{\hat H}_{PQ}&{\hat H}_{PQ}&\ldots&{\hat H}_{PP}\cr}
\pmatrix{{\gamma}_1{\hat Q}|{\phi}_1{\rangle}\cr
{\gamma}_2{\hat Q}|{\phi}_2{\rangle}\cr
\vdots\cr
{\hat P}|E{\rangle}\cr}=E\pmatrix{{\gamma}_1{\hat Q}|{\phi}_1{\rangle}\cr
{\gamma}_2{\hat Q}|{\phi}_2{\rangle}\cr
\vdots\cr
{\hat P}|E{\rangle}\cr}\eqno(3.10)$$
This yields a series of equations
$${\hat H}_{QQ}|{\phi}_j{\rangle}{\gamma}_j+{\hat H}_{QP}|E{\rangle}=
E{\hat Q}|{\phi}_j{\rangle}{\gamma}_j,\eqno(3.11)$$
for $j=1,2,...$ and
$${\hat H}_{PP}|E{\rangle}+{\sum_j}H_{PQ}|{\phi}_j{\rangle}{\gamma}_j
=E{\hat P}|E{\rangle}.\eqno(3.12)$$
Before we can proceed further, we must find conditions for Hermiticity of
the Hamiltonian,
${\hat H}$.
\bigskip
\section{Hermiticity Condition}

Consider the arbitrary states, $|{\Psi}{\rangle}$ and $|{\Xi}{\rangle}$.
The condition for
Hermiticity of these states is that
$${\langle}\Xi|{\hat H}|\Psi{\rangle}-{\langle}\Psi|{\hat
H}|\Xi{\rangle}^*=0.\eqno(4.1)$$
We can decompose the states $|{\Psi}{\rangle}$ and $|{\Xi}{\rangle}$
into their contributions to the two disjoint configuration space
regions and write them
in the form,
$$|{\Psi}{\rangle}={\hat Q}|{\Psi}{\rangle}+{\hat
P}|{\Psi}{\rangle}={\sum_j}{\beta}_j{\hat
Q}|{\phi}_j{\rangle} +{\hat P}|{\Psi}{\rangle},\eqno(4.2)$$
where ${\beta}_j={\langle}{\phi}_j|{\hat Q}|{\Psi}{\rangle}$ and
$$|{\Xi}{\rangle}={\hat Q}|{\Xi}{\rangle}+{\hat
P}|{\Xi}{\rangle}={\sum_j}{\alpha}_j{\hat Q}|{\phi}_j{\rangle}
+{\hat P}|{\Xi}{\rangle},\eqno(4.3)$$
where ${\alpha}_j={\langle}{\phi}_j|{\hat Q}|{\Xi}{\rangle}$. In the
interior region, we have
expanded $|{\Psi}{\rangle}$ and $|{\Xi}{\rangle}$ in terms of the complete
set of energy
eigenstates, ${\hat Q}|{\phi}_j{\rangle} $,  of the Hamiltonian,
${\hat H}_{QQ}$.
To simplify the notation, let
$|{\psi}_j{\rangle}{\equiv}{\beta}_j{\hat Q}|{\phi}_j{\rangle}$,
$|{\xi}_j{\rangle}{\equiv}{\alpha}_j{\hat Q}|{\phi}_j{\rangle}$,
${\hat P}|{\Psi}{\rangle}=|{\Psi}_P{\rangle}$, and
${\hat P}|{\Xi}{\rangle}=|{\Xi}_P{\rangle}$.
 The states can be written
$$|{\Xi}{\rangle}=\pmatrix{|{\xi}_1{\rangle}\cr
|{\xi}_2{\rangle}\cr
\vdots\cr
|{\Xi}_P{\rangle}\cr} ~~{\rm and}~~
|{\Psi}{\rangle}=\pmatrix{|{\psi}_1{\rangle}\cr
|{\psi}_2{\rangle}\cr
\vdots\cr
|{\Psi}_P{\rangle}\cr}.\eqno(4.4)$$

If we note that
$${\langle}\Xi|{\hat H}|\Psi{\rangle}=({\langle}{\xi}_1|,{\langle}{\xi}_2|,...
{\langle}{\Xi}_P|)\pmatrix{{\hat H}_{QQ}&0&\ldots&{\hat H}_{QP}\cr
0&{\hat H}_{QQ}&\ldots&{\hat H}_{QP}\cr
\vdots&\vdots&\ddots&\vdots\cr
{\hat H}_{PQ}&{\hat H}_{PQ}&\ldots&{\hat H}_{PP}\cr}
\pmatrix{|{\psi}_1{\rangle}\cr
|{\psi}_2{\rangle}\cr
\vdots\cr
|{\Psi}_P{\rangle}\cr}$$
$$~~~={\sum_j}{\langle}{\xi}_j|{\hat H}_{QQ}|{\psi}_j{\rangle}+
{\sum_j}{\langle}{\xi}_j|{\hat H}_{QP}|{\Psi}_P{\rangle}+
{\sum_j}{\langle}{\Xi}_P|{\hat H}_{PQ}|{\psi}_j{\rangle}+
{\langle}{\Xi}_P|{\hat H}_{PP}|{\Psi}_P{\rangle},\eqno(4.5)$$
then the condition for Hermiticity is
$${\langle}\Xi|{\hat H}|\Psi{\rangle}-{\langle}\Psi|{\hat H}|\Xi{\rangle}^*=
{\sum_j}[{\langle}{\xi}_j|{\hat H}_{QQ}|{\psi}_j{\rangle}
-{\langle}{\psi}_j|{\hat H}_{QQ}|{\xi}_j{\rangle}^*]$$
$$~~~+
{\sum_j}[{\langle}{\xi}_j|{\hat H}_{QP}|{\Psi}_P{\rangle}
-{\langle}{\psi}_j|{\hat H}_{QP}|{\Xi}_P{\rangle}^*]+
{\sum_j}[{\langle}{\Xi}_P|{\hat H}_{PQ}|{\psi}_j{\rangle}
-{\langle}{\Psi}_P|{\hat H}_{PQ}|{\xi}_j{\rangle}^*]$$
$$~~~+[{\langle}{\Xi}_P|{\hat H}_{PP}|{\Psi}_P{\rangle}
-{\langle}{\Psi}_P|{\hat H}_{PP}|{\Xi}_P{\rangle}^*]=0.\eqno(4.6)$$
We can now evaluate Eq. (4.6) term by term.  Let
${\langle}x|{\psi}_j{\rangle}{\equiv}{\psi}_j(x)$,
${\langle}x|{\xi}_j{\rangle}{\equiv}{\xi}_j(x)$,
${\langle}x|\Xi_P{\rangle}{\equiv}{\Xi}_P(x)$
and ${\langle}x|{\Psi}_P{\rangle}{\equiv}{\Psi}_P(x)$. Then
$${\langle}\Xi|{\hat H}|\Psi{\rangle}-{\langle}\Psi|{\hat H}|\Xi{\rangle}^*=
-{{\hbar}^2\over 2m}{\sum_j}{\biggl(}{\xi}_j^*(a){d{\psi}_j\over dx}{\bigg|}_a-
{\psi}_j^*(a){d{\xi}_j\over dx}{\bigg|}_a{\biggr)}$$
$$~~~+{C\over 4}{\sum_j}{\biggl(}{\xi}_j^*(a){d{\Psi}_P\over dx}{\bigg|}_a-
{\psi}_j^*(a){d{\Xi}_P\over dx}{\bigg|}_a{\biggr)}+
{C\over 4}{\sum_j}{\biggl(}{\Xi}_P^*(a){d{\psi}_j\over dx}{\bigg|}_a-
{\Psi}_P^*(a){d{\xi}_j\over dx}{\bigg|}_a{\biggr)}$$
$$~~~-{{\hbar}^2\over 2m}{\biggl(}{\Xi}_P^*(a){d{\Psi}_P\over dx}{\bigg|}_a-
{\Psi}_P^*(a){d{\Xi}_P\over dx}{\bigg|}_a{\biggr)}=0,\eqno(4.7)$$
is the Hermiticity condition.
For our subsequent discussion, it is useful to note that for the special
case when the
eigenstate, ${\phi}_j(x)$, has zero slope on the interface,
 so ${d{\phi}_j\over dx}{\big|}_a=0$, the Hermiticity condition reduces to
$${\langle}\Xi|{\hat H}|\Psi{\rangle}-{\langle}\Psi|{\hat H}|\Xi{\rangle}^*=
{C\over 4}{\sum_j}{\biggl(}{\xi}_j^*(a){d{\Psi}_P\over dx}{\bigg|}_a-
{\psi}_j^*(a){d{\Xi}_P\over dx}{\bigg|}_a{\biggr)}$$
$$~~~-{{\hbar}^2\over 2m}{\biggl(}{\Xi}_P^*(a){d{\Psi}_P\over dx}{\bigg|}_a-
{\Psi}_P^*(a){d{\Xi}_P\over dx}{\bigg|}_a{\biggr)}=0,\eqno(4.8)$$

If we let $|{\Psi}{\rangle}=|E{\rangle}$ and $|{\Xi}{\rangle}=|E{\rangle}$
in Eq. (4.8), the
Hermiticity condition reduces to
$${C\over 4}{\sum_j}{\gamma}_j{\phi}_j(a)={{\hbar}^2\over
2m}{\Psi}_{E,P}(a),\eqno(4.9)$$
where ${\Psi}_{E,P}(x)={\langle}x|{\hat P}|E{\rangle}$.
This Hermiticity condition allows us to determine the coupling constant, $C$.
If we remember that ${\langle}x|{\hat
Q}|E{\rangle}={\sum_j}{\gamma}_j{\phi}_j(x)$, then continuity of the
energy eigenfunction at
$x=a$ requires that
$$C={2{\hbar}^2\over m}.\eqno(4.10)$$
Thus, the Hermiticity condition allows us to determine the strength of the
coupling at the
interface.

\bigskip
\section{The Reaction Function}

We can use the formalism derived in Sections (3) and (4) to derive
an expression for the reaction function, $R(E)$.
Let us return to Eq. (3.11).  If we multiply by ${\langle}{\phi}_j|{\hat
Q}$,
use Eq. (3.4), and the normalization condition, ${\langle}{\phi}_j|{\hat
Q}|{\phi}_j{\rangle}=1$, we obtain
$$({\lambda}_j-E){\gamma}_j+{C\over 4}{\phi}_j^*(a){d{\Psi}_R^0\over
dx}{\bigg|}_a=0.\eqno(5.1)$$
Similarly, if we multiply  Eq. (3.12) by ${\langle}E|{\hat P}$, we obtain
$$-{{\hbar}^2\over
2m}{\int_a^{\infty}}dx~{\Psi}_{E,P}^{*}(x){d^2{\Psi}_{E,P}\over dx^2}
+{C\over 4}{\sum_j}{\Psi}_{E,P}^{*}(a){d{\phi}_j\over dx}{\bigg|}_a{\gamma}_j$$
$$~~~=E{\int_a^{\infty}}dx~{\Psi}_{E,P}^{*}(x){\Psi}_{E,P}(x),\eqno(5.2)$$
where ${\Psi}_{E,P}(x){\equiv}{\langle}x|{\hat P}|E{\rangle}$.

Before we can proceed further, we must decide on boundary conditions for
our states, ${\phi}_j(x)$.  We can use any complete basis set in the region,
$0<x<a$, provided they have the property ${\phi}_j(0)=0$. We are free to choose
the boundary condition at $x=a$. We have seen in Eq. (4.8) that the Hermiticity
condition becomes especially simple for the boundary condition,
${d{\phi}_j(x)\over dx}{\bigg|}_a=0$. Therefore, that is the boundary condition
that we will use here.
The states with these boundary conditions are given by
${\phi}_j(x)={\sqrt{2\over a}}{\sin}{\bigl(}{j{\pi}x\over 2a}{\bigr)}$,
for $j$ odd $(j=1,3,...)$, and the corresponding eigenvalues are
${\lambda}_j={{\hbar}^2{\pi}^2j^2\over 8a^2m}+V_0$. The states,
${\phi}_j(x)$ and
the eigenvalues, ${\lambda}_j$, should not be confused with the exact energy
eigenvalues and eigenstates in Region I. The states, ${\phi}_j(x)$,
only serve to provide us a complete set of states in Region I which can then
be used to construct the exact energy eigenstates in Region I.
With these boundary conditions, Eq. (5.2) takes the form
$$-{{\hbar}^2\over
2m}{\int_a^{\infty}}dx~{\Psi}_{E,P}^{*}(x){d^2{\Psi}_{E,P}\over dx^2}
=E{\int_a^{\infty}}dx~{\Psi}_{E,P}^{*}(x){\Psi}_{E,P}(x).\eqno(5.3)$$

Let us now show that this description of the system leads to
well known
results. We can derive the reaction function originally obtained by Wigner
and Eisenbud.
Let us return to Eq. (5.1) and solve for ${\gamma}_j$. We find
$${\gamma}_j={{\hbar}^2\over 2m}{1\over E-{\lambda}_j}{\phi}_j^*(a)
{d{\Psi}_{E,P}\over dx}{\bigg|}_a.\eqno(5.4)$$
If we combine Eqs. (4.9) and (5.4), we obtain
$${\Psi}_{E,P}(a)={{\hbar}^2\over 2m}{\sum_j}{{\phi}_j^*(a){\phi}_j(a)\over
E-{\lambda}_j}
{d{\Psi}_{E,P}\over dx}{\bigg|}_a.\eqno(5.5)$$
From Eq. (5.5), we can express the logarithmic derivative
of the outside wavefunction,
${d{\Psi}_E^0\over dx}{a\over {\Psi}_{E,P}}$, evaluated at the interface
(and therefore the reaction function, $R(E)$),
in terms of the inside states. We find
$$R(E){\equiv}{{\Psi}_{E,P}(a)\over a{d{\Psi}_{E,P}\over dx}{\bigg|}_a}
={{\hbar}^2\over 2ma}{\sum_j}{{\phi}_j^*(a){\phi}_j(a)\over
E-{\lambda}_j}.\eqno(5.6)$$
For scattering
in higher dimensional space or for systems with internal degrees of freedom
the reaction function becomes a matrix and is then called the {\it reaction
matrix}.

In Eq. (2.8), we obtained an exact expression for the reaction function. If
we equate
Eqs. (2.8) and (5.6), we obtain
$$R(E)={{\tan}(k'a)\over k'a}={{\hbar}^2\over
2ma}{\sum_j}{{\phi}_j^*(a){\phi}_j(a)\over
E-{\lambda}_j}= {{\hbar}^2\over ma^2}{\sum_{j=1}^{\infty}}
{{\sin}^2{\bigl(}{(2j-1){\pi}\over 2}{\bigr)}\over
E-{{\hbar}^2{\pi}^2(2j-1)^2\over 8a^2m}-V_0},
\eqno(5.7)$$
for $j$ odd.
Let ${\kappa}^2={2mV_0\over {\hbar}^2}$. Then we can write Eq. (5.7) in the form
$${{\tan}(\sqrt{k^2-{\kappa}^2}a)\over \sqrt{k^2-{\kappa}^2}a}=
 {2\over a^2}{\sum_{\matrix{j=1\cr odd\cr}}^{\infty}}
{1\over  k^2-{\kappa}^2-{{\pi}^2(2j-1)^2\over 4a^2}}
\eqno(5.8)$$
From Ref. [\cite{grad}] one can show that this is just
the definition of the series expansion of the tangent function in terms of its
argument.

The reaction function can also be expressed as a product of matrices.
Let us consider the first $N$
eigenvalues and eigenvectors of the the Hamiltonian, ${\hat H}_{QQ}$ (we
later let
$N{\rightarrow}\infty$). We then write the Hamiltonian, ${\hat H}_{QQ}$,
in the matrix representation in which it is diagonal (in this representation
we denote it as
${\bar H}_{in}^N$), and obtain
$${\bar H}_{in}^N=
\pmatrix{{\lambda}_1&0&\ldots&0\cr
0&{\lambda}_2&\ldots&0\cr
\vdots&\vdots&\ddots&\vdots\cr
0&0&\ldots&{\lambda}_N\cr}.\eqno(5.9)$$
We also introduce the vector of eigenstates of ${\hat H}_{QQ}$, but
evaluated at the interface,
$${\bar v}_N=\sqrt{{{\hbar}^2\over 2ma}}
\pmatrix{{\phi}_1(a)\cr
{\phi}_2(a)\cr
\vdots\cr
{\phi}_N(a)\cr}.\eqno(5.10)$$
Then the reaction function can be written in the compact form
$$R_N(E)={\bar v}_N^{\dagger}{1\over E{\bar 1}_N-{\bar H}^N_{in}}{\bar v}_N
={{\hbar}^2\over 2ma}{\sum_j^N}{{\phi}_j^*(a){\phi}_j(a)\over
E-{\lambda}_j},
\eqno(5.11)$$
where ${\bar 1}_N$ is the $N{\times}N$ dimensional unit matrix and
${\bar v}_N^{\dagger}$ denotes
the Hermitian adjoint of ${\bar v}_N$.

\section{ The Scattering Function}

It is useful to express the scattering function, $S(E)$, in matrix notation,
because we can then find a very interesting expression for the complex
energy poles of $S(E)$. In Eq. (2.7), we obtained an expression for the
scattering function in terms
of the reaction function. In terms of the truncated reaction function it is
$$S_N(E)={\rm e}^{-2ika}{\biggl[}{1+ikaR_N(E)\over
1-ikaR_N(E)}{\biggr]},\eqno(6.1)$$
where
$$R_N(E)={{\hbar}^2\over ma^2}{\sum_{j=1}^{N}}
{{\sin}^2{\bigl(}{(2j-1){\pi}\over 2}{\bigr)}\over
E-{{\hbar}^2{\pi}^2(2j-1)^2\over 8a^2m}-V_0}.
\eqno(6.2)$$
In Table (1), we give the positions of the three lowest energy quasi-bound state
poles found by using the exact expression for $R(E)$ given in Eq. (2.8). These
values are exact to the number of digits shown. Also in Table (1), we give the
positions of those poles found by using Eq. (6.1) for four different
truncations,
$N=250$, $N=350$, $N=500$, and $N=1000$. The values obtained using Eq.
(6.1) converge
slowly to the correct answer.

\begin{table}
\caption{comparison of pole positions with exact result}
\begin{tabular}{||l|l|l|l|l||}  \hline
n       &  First Pole       &  Second Pole      &  Third Pole       \\
\hline
250     &  17.9812   -5.0071 &  45.0729  -17.6046 &  93.2048  -34.5167 \\
\hline
350     &  17.9771   -4.9999 &  45.0197  -17.5295 &  92.8700  -34.2099 \\
\hline
500     &  17.9742   -4.9945 &  44.9806  -17.4736 &  92.6336  -33.9798 \\
\hline
1000    &  17.9707   -4.9883 &  44.9359  -17.4089 &  92.3725  -33.7124 \\
\hline
exact   &  17.9672   -4.9821 &  44.8921  -17.3448 &  92.1262  -33.4466 \\
\hline
\end{tabular}
\end{table}

The scattering function  can be written in several other forms as well.
Let us introduce the column vector,
$${\bar w}_N=\sqrt{ka}{\bar v}_N=\sqrt{{{\hbar}^2k\over 2m}}
\pmatrix{{\phi}_1(a)\cr
{\phi}_2(a)\cr
\vdots\cr
{\phi}_N(a)\cr}.\eqno(6.3)$$
Then the scattering function becomes
$$S_N(E)={\rm e}^{-2ika} {\biggl[}{{1+iK_N}\over {1-iK_N}}{\biggr]}
={\rm e}^{-2ika} {\biggl[}1+{{2iK_N}\over {1-iK_N}}{\biggr]}    .\eqno(6.4)$$
where
$$K_N{\equiv}{\bar w}_N^{\dagger}{1\over E{\bar 1}_N-{\bar
H}^N_{in}}{\bar w}_N.\eqno(6.5)$$
Using Eq. (6.4), we can write (we suppress the index $N$)
$${K\over 1-iK}={1\over 1-i{\bar w}^{\dagger}{1\over E{\bar 1}-{\bar
H}_{in}}{\bar w}}
{\bar w}^{\dagger}{1\over E{\bar 1}-{\bar H}_{in}}{\bar w}$$
$$~~~={\sum_{n=0}^{\infty}}(i)^n{\bar w}^{\dagger}
{\biggl(}{1\over E{\bar 1}-{\bar H}_{in}}
{\bar w}{\bar w}^{\dagger}{\biggr)}^n
{1\over E{\bar 1}-{\bar H}_{in}}{\bar w})$$
$$~~~={\bar w}^{\dagger}( E{\bar 1}-{\bar H}_{in})
{\biggl(}{1\over E{\bar 1}-{\bar H}_{in}-i{\bar w}{\bar
w}^{\dagger}}{\biggr)}{1\over  E{\bar 1}-{\bar H}_{in}}{\bar w},\eqno(6.6)$$
The scattering function, $S_N(E)$, can now be written in the form
$$S_N(E)=1+2i{\bar w}_N^{\dagger}( E{\bar 1}_N-{\bar H}^N_{in})
{{\bar M}\over {\rm Det}[E{\bar 1}_N-{\bar H}^N_{in}-i{\bar w}_N{\bar
w}_N^{\dagger}]}{1\over  E{\bar 1}_N-{\bar H}^N_{in}}{\bar w}_N.\eqno(6.7)$$
where ${\bar M}$ is the adjoint of the matrix,
$E{\bar 1}-{\bar H}_{in}-i{\bar w}{\bar w}^{\dagger}$, and ${\rm Det}$ denotes
its determinant.

The poles of the scattering function are given by the condition
$${\rm Det}[E{\bar 1}_N-{\bar H}^N_{in}-i{\bar w}_N{\bar w}_N^{\dagger}]=0.
\eqno(6.8)$$
Eq. (6.8) gives an $N^{th}$ order polynomial in $E$ whose solutions are the
complex energies which locate the poles.
At first sight, it would appear that the complex energy poles of the
scattering function  are simply
given by the eigenvalues of the the non-Hermitian matrix,
${\bar H}^N_{eff}{\equiv}{\bar H}^N_{in}+i{\bar w}_N{\bar
w}_N^{\dagger}$. However,
the column matrices,
${\bar w}_N$ depend on energy, $E$, and therefore the eigenvalues,
 ${\mu}_i(E)$,
 of   ${\bar H}^N_{eff}$ also depend on energy, $E$.

In Fig. (5), we locate the
zeros of
${\rm Det}[E{\bar 1}-{\bar H}_{eff}]=
{\rm Det}[E{\bar 1}-{\bar H}_{eff}-i{\bar w}{\bar w}^{\dagger}]$
in the neighborhood of the first three resonance energies.
The zeros satisfy the equation,
$${\rm Det}[E{\bar 1}-{\bar H}_{eff}]=(E-{\mu}_1(E))(E-{\mu}_2(E))
{\times}...{\times}(E-{\mu}_N(E)).\eqno(6.9)$$
In Fig. (5.a) we plot $(E-{\mu}_1(E))$ versus $E$ for complex values of $E$
in the
neighborhood of the first pole. We see that $(E-{\mu}_1(E))=0$ at the energy of
the first pole. In Figs. (5.b) and (5.c), we plot
$(E-{\mu}_1(E))$ and $(E-{\mu}_1(E))$ at the energies of the second and
third poles, respectively. Again we see that they go to zero at the
energies of their
respective poles.

\section{Random potential}

The potential used in Section (2), is a smooth step function which allows
us to solve the scattering problem exactly. It also allows us to obtain exact
expressions for the basis state energies, $\lambda_j$  and 
coupling constants, ${\bar v}_N$. For our simple scattering problem, matrix
elements of ${\bar v}_N$ simply alternate between two constant values. 

We now use these same methods to study scattering from a random step
potential.  The random potential we use is a sequence of 10 tent-like shapes
(upright or inverted) on the interval $0<x<a$ with $a=100$. We again choose
${\hbar}=1$ and $m=1$.  We can express the potential in the form
$$
V(x)=10+\sum_{j=1}^{10} \left ( \frac{(v_j-v_{j-1})}{10}(x-10j+10)
+v_{j-1}\right ) \Theta (10j-x) \Theta (x-10j+10)
\eqno(7.1)
$$
The values of $v_j$ are chosen at random from a uniform distribution 
over the interval,
$-0.5{\leq}v_j{\leq}+0.5$.
$\Theta(x)$ is the heaviside function. 
One hundred different realizations of this potential are shown in Fig. 6. 

For this larger value of $a$, there are  about 10 to 12
scattering resonance peaks in the  energy interval $10.5{\leq}E{\leq}11$. To
find the eigenvalues and eigenfunctions  for this random potential we used an
implicit finite difference method. When there is no randomness ($v_j=0$ for
all $j$), the Wigner delay time can be calculated exactly as we showed in
Section 2. Fig. (7) shows the Wigner delay times when $v_j=0$ for
all $j$. In  Figure (7) , the exact result is given by the solid line, and 
the approximate result (obtained by using the reaction matrix series
expansion, Eq. (5.6)) is given by the discrete dots. The agreement is
excellent. In Fig. (8), we plot the values of
$v_N$ which are obtained from the 100 hundred realizations of the random
potential when 
$a=100$. The peak points (positive and negative) on the oscillating solid
line give values of $v_N$ for the case
$v_j=0$ for all
$j$.  We see that the higher eigenmodes are not
affected by the  random potential.

In Fig. (9), we show the  Wigner delay times for 100 different realizations
of random potential. The positions of resonance peaks as well as their widths
change with different potentials.  In Fig. (10) we plot a histogram of
values of the Wigner  delay times shown in Figure (9). This is similar to
distribution one can get from RMT calculations, except that our
distribution has a longer tail (see
\cite{kn:10}~\cite{seba}).  Finally, in Fig. (11)  we show the distribution
of Wigner delay time resonance widths.   We find
that the distribution of delay time widths is fairly symmetrically
distributed around its average value.  

As we can see, this approach allows us to compute the scattering properties
of a variety of shapes of scattering potential with great accuracy. The only
constraint is that the scattering potential must occupy a well defined
region of space. 

\pagebreak

\newpage

\section{Conclusions}

The configuration space scattering picture, described in Sections (3)-(6)
 is a basis for Weidenmuller's phenomenological Hamiltonian theory of nuclear
scattering, and it has provided a framework to look for universal behavior
 in the scattering properties of
open quantum systems whose underlying classical counterparts are
 chaotic [\cite{kn:8}].
There is now a large body of work [\cite{kn:7}]
 which shows that bounded quantum systems,
whose underlying classical dynamics is chaotic, have energy eigenvalues and
energy eigenfunction which have statistical properties similar to those of
certain types of random Hamiltonian matrices.
Similar questions are being asked about the scattering properties of open
quantum systems
For open systems, one can look at the
statistical properties of the spacings and widths of scattering resonances,
and of
Wigner delay times.

The theory described in Sections (3)-(6) can be used to compare
the results of laboratory and numerical experiments to random Hamiltonian
matrix
predictions. For example, for scattering systems in which no magnetic fields are
present, the Hamiltonian, $H_{in}$, which is formed from the eigenvalues of
$H_{QQ}$,
is replaced by a Hamiltonian, $H_{in}^{rm}$, which is formed from the
eigenvalues
of a Hamiltonian matrix whose matrix elements are random numbers which are
Gaussian
distributed. The elements of the coupling matrix, ${\bar w}$, are also chosen
from a random distribution. The strength of the coupling constant, $C$, is left
as a variable parameter in these random matrix theories. The random matrix
theory predictions are in qualitative agreement with numerical experiments and
some laboratory experiments.

However, a  number of assumptions underly
the random matrix theory predictions, and there are a number of issues that
remain open. For actual scattering systems, the coupling constant, $C$,
is fixed by the detailed dynamics using the Hermiticity condition, but
in random matrix theories it is a variable parameter.
  In random matrix theories, the
coupling matrix, ${\bar w}$, is chosen from a random distribution. How does
it actually look
for a deterministic system with underlying classical chaos? This has never
been studied.

Random matrix theories
also neglect the dependence of $H_{eff}$ on the energy, $E$.  The actual
error made
in making this assumption needs to be understood.  For two dimensional
electron waveguides, there
is an additional problem that at the threshold energy where a new channel
opens,
quasi-bound states inside the cavity can extend far down the leads.~\cite{akg}
 It is
not clear
how well the theory presented in Sections (3)-(6) and therefore the random
matrix
theory,  can describe scattering in
those regions.  Never-the-less, this approach to scattering theory has allowed
contact to be made between chaos theory and scattering theory for open
quantum systems. As we have just shown, it can also provide a powerful tool
to study scattering processes in atomic and mesoscopic devices, when
 intrinsic disorder in the medium needs to be included.

\section{Acknowledgements}

 The authors wish to thank the Robert A. Welch Foundation Grant No. F-1051,
 NSF Grant No. INT-9602971, and DOE Contract No. DE-FG03-94ER14405 for
 partial support of this work. We thank NPACI and the University of Texas
at Austin
 High Performance Computing Center for use of their computer facilities.
 We also thank German Luna Acosta and Gengiz Sen for helpful
discussions.

 \pagebreak

\pagebreak
\newpage

\begin{figure}
\caption{The scattering potential, $V(x)$ for ${\hbar}=1$, $V_0=10$, and $a=1$.
 }
\label{figure1}
\end{figure}

\begin{figure}
\caption{
(a) Phase $\theta$ of $S=exp(i\theta)$ function,
(b) Wigner delay time $\tau$  
Both plot is as a function of
 energy for ${\hbar}=1$, $V_0=10$,
and $a=1$. Three resonances are
shown with arrows at the resonance energy values.}
\label{figure2}
\end{figure}

\begin{figure}
\caption{Absolute value of exact energy eigenstates plotted as a function of
spatial variable, $x$, and energy, $E$, for ${\hbar}=1$, $V_0=10$,
and $a=1$.}
\label{figure3}
\end{figure}

\begin{figure}
\caption{Poles of S function in the lower complex plane for ${\hbar}=1$,
$V_0=10$,
and $a=1$. The absolute value
of S is plotted in logarithmic scale, as a function  to real and imaginary
parts of the energy.}
\label{figure4}
\end{figure}

\begin{figure}
\caption{
(a) $E-{\mu}_1(E)$ versus $E=E_r+iE_i$ in the neighborhood of the first pole;
(b) $E-{\mu}_2(E)$ versus $E=E_r+iE_i$ in the neighborhood of the second pole;
(c) $E-{\mu}_3(E)$ versus $E=E_r+iE_i$ in the neighborhood of the third pole.
All plots are for ${\hbar}=1$, $V_0=10$,  and $a=1$.}
\label{figure5}
\end{figure}

\begin{figure}
\caption{
One hundred realizations of the random potential, Eq. (7.1).}
\label{figure6}
\end{figure}

\begin{figure}
\caption{
Wigner delay time for the step potential, $V(x)=10$ for $0<x<100$ and
$V(x)=0$ for $x>100$. The solid line is the exact result. The dots are
obtained using the series approximation to the reactions matrix, Eq. (5.6)}
\label{figure7}
\end{figure}

\begin{figure}
\caption{
The coupling constants, $v_N$, for the 100  different realizations of random
potential shown in Fig. (6). The peak values (positive and negative) of the
solid line give the values of $v_N$ for $v_j=0$ and $V(x)=10$.}
\label{figure8}
\end{figure}

\begin{figure}
\caption{
Wigner delay times for the 100 realizations of the random potential shown in 
Fig.~6. }
\label{figure9}
\end{figure}

\begin{figure}
\caption{
The normalized probability distribution, $P(\tau)$, of scaled Wigner delay
times, ${\tau}/{\langle}\tau{\rangle}$, where ${\langle}\tau{\rangle}=125.42$
is the average value of $\tau$, taken over all 100 realizations of the random
potential. } 
\label{figure10}
\end{figure}

\begin{figure}
\caption{
The normalized probability distribution, $P({\Gamma})$, of scaled Wigner
delay time half widths, ${\Gamma}/{\langle}\Gamma{\rangle}$, where 
${\langle}\Gamma{\rangle}=0.05$ is average half-width taken over all
100 realizations of the random potential.}
\label{figure11}
\end{figure}

\end{document}